\documentclass[aps,pra,twocolumn,superscriptaddress,nofootinbib]{revtex4-2}

\usepackage[braket, qm]{qcircuit} 
\usepackage{graphicx}
\usepackage{amssymb}
\usepackage{indentfirst}
\usepackage{amsmath}
\usepackage{color,xcolor}
\usepackage{subfigure}
\usepackage{verbatim}
\usepackage[noend]{algpseudocode}
\usepackage{algorithm,algorithmicx}
\usepackage{amsthm}

\usepackage{tikz}

\begin{document}

\title{Determination of all unknown pure quantum states with two observables}

\author{Yu Wang}
\email[]{ming-jing-happy@163.com}
\affiliation{Beijing Institute of Mathematical Sciences and Applications, Beijing 101408, China}

\begin{abstract}
Efficiently extracting information from pure quantum states using minimal observables on the main system is a longstanding and fundamental issue in quantum information theory. Despite the inability of probability distributions of position and momentum to uniquely specify a wavefunction, Peres conjectured a discrete version wherein two complementary observables, analogous to position and momentum and realized as projective measurements onto orthogonal bases, can determine all pure qudits up to a finite set of ambiguities. Subsequent findings revealed the impossibility of uniquely determining $d$-dimenisonal pure states even when neglecting a measure-zero set with any two orthogonal bases, and Peres's conjecture is also correct for \(d=3\) but not for \(d=4\). 
In this study, we show that two orthogonal bases are capable of effectively filtering up to \(2^{d-1}\) finite candidates by disregarding a measure-zero set, without involving complex numbers in the bases' coefficients. Additionally, drawing inspiration from sequential measurements to directly calculate the target coefficients of the wavefunction using two complementary observables, we show that almost all pure qudits can be uniquely determined by adaptively incorporating a POVM in the middle, followed by measuring the complementary observable.

\end{abstract}

\maketitle

According to the postulates of quantum mechanics, the quantum states of a closed physical system are represented by unit vectors in a Hilbert space \cite{nielsen2010}. When measuring a pure state \(|\psi\rangle=\sum_{k=0}^{d-1} a_k|k\rangle\) using an observable \(O=\sum_{k=0}^{d-1}\lambda _k |\psi_k\rangle\langle \psi_k|\) in a finite-dimensional Hilbert space \(\mathcal{H}_d\), we can obtain the outcome \(\lambda_k\) with a probability \(|\langle \psi|\psi_k\rangle|^2\) by Born rule \cite{landau2013quantum}. With each observable, the probability distribution \(\{|\langle \psi|\psi_k\rangle|^2\}\) encapsulates the original information of the quantum state \(|\psi\rangle\).

A fundamental question arises: How can we extract as much information as possible about \(|\psi\rangle\) using the fewest possible number (two) of observables?

This problem finds its roots in Pauli's conjecture. In classical physics, possessing complete knowledge of both the position and momentum of a state enables us to infer and uniquely determine the underlying mechanism. The position and momentum can be measured simultaneously with infinite accuracy \cite{kibble2004classical}. When extended to the quantum realm, Pauli's conjecture concerns whether the probability distributions of position and momentum can determine (up to a global phase) all wavefunctions \cite{pauli1933handbuch}. However, the concept of Pauli nonunique arises from the existence of pairs of distinct wavefunctions that yield identical position and momentum probability distributions \cite{corbett1977wave,vogt1978position,pavivcic1987complex,busch1989determination,weigert1992pauli,weigert1996determine}. Asher Peres investigated a specific finite version of this problem in his famous textbook \cite{peres1997}. 
The two complementary observables can be likened to projective measurements onto two orthonormal bases \(\mathcal{B}_0=\{|0\rangle,\cdots,|d-1\rangle\}\) and \(\mathcal{B}_1=F\mathcal{B}_0\), where \(F\) represents the quantum Fourier transformation. 
Peres thought that all pure quantum states $|\psi\rangle$ should be determined by the two probability distributions $\{|\langle \psi  |k\rangle|^2\}$  and $\{|\langle \psi  |F|k\rangle|^2\}$, up to a finite set of ambiguities.

A series of research studies demonstrate the impossibility of using two orthonormal bases for the unique determination of pure states. Even when neglecting a failure set limited to measure-zero compared to the entire pure state space, Flammia, Silberfarband, and Caves proved that not only the two complementary bases, but any two orthonormal bases, are insufficient for unique determination \cite{flammia2005}. Theoretically, for dimensions \(d=3\) and \(d\geq 5\), the minimal number of orthonormal bases required for unique determination is four \cite{heinosaari2013,mondragon2013,jaming2014,carmeli2015many}. 
The remaining unresolved question is whether there exist minimal three orthonormal bases capable of uniquely determining all pure states for dimension \(d=4\). Considering the original Peres's conjecture, it is shown to be correct for dimension \(d=3\), as at most six candidates match the probability distributions of Peres's bases \(\mathcal{B}_0\) and \(\mathcal{B}_1\). However, for \(d=4\), there are infinitely many candidates \cite{sun2020}.

For unknown $d$-dimensional pure states $|\psi\rangle$, except for a measure-zero set (almost all pure states), a series of works has focused on their unique determination using various projective measurement resources. Here, we briefly mention a few. These include employing $3d-2$ rank-1 projections \cite{finkelstein2004pure,wang2018pure}, constructing five orthonormal bases \cite{goyeneche2015five}, utilizing two or three orthonormal bases with an ancilla qubit \cite{wang2022pure}, and exploring the use of three orthonormal bases \cite{zambrano2024minimal}. 
For $n$-qubit systems with dimension $d=2^n$, the consideration extends to constructing $mn+1$ separable orthonormal bases with $m\geq 2$ \cite{pereira2022scalable}, as well as $2n+1$ eigenbases of special Pauli observables \cite{verdeil2023pure}. With these methods, the number of measurement outcomes decreases drastically compared to the conventional measurement resource of $3^n$ Pauli observables, or randomized $O(2^n n^2)$ Pauli observables by compressed sensing \cite{gross2010quantum}. Without the limitation of unique determination, three orthonormal bases are constructed to filter at most $2^{d-1}$ candidates for almost all pure states \cite{zambrano2020estimation}.   

Complementary observables demonstrate their potential to directly obtain information about pure quantum states with an ancilla coupling pointer. The complex value of $\psi(x,t)$ at each position $x$ can be directly obtained by weakly measuring position, followed by a strong measurement of momentum \cite{lundeen2011direct}. Subsequently, many schemes have been developed utilizing the framework of sequential measurements to directly acquire pertinent information about unknown quantum states \cite{vallone2016strong,malik2014direct,salvail2013full,mirhosseini2014compressive,piacentini2016measuring,pan2019direct}. 

In this study, we explore the determination of unknown pure states using two observables. Firstly, we transform the basis $\mathcal{B}_1$ into another types of orthonormal basis $\mathcal{C}_1=U\mathcal{B}_0$. Theoretically, we demonstrate that for almost all pure states $|\psi\rangle$, disregarding a measure-zero set, two probability distributions $\{|\langle \psi |k\rangle|^2\}$ and $\{|\langle \psi |U|k\rangle|^2\}$ can generate a finite set of $2^{d-1}$ candidates. Notably, the unitary operation \(U\), where all elements below the secondary diagonal are set to zero, can contain no complex numbers, typically deemed essential for quantum explanations \cite{renou2021quantum}. Secondly, we categorize all pure states \(|\psi\rangle\) into \(2^d-1\) classes based on their amplitudes. Through adaptive POVM followed by a complementary observable \(\mathcal{B}_1\), almost all classes of pure states can be uniquely determined.

\section{Two orthonormal bases for filtering almost all pure  qudits into a finite set}

In a \( d \)-dimensional Hilbert space, any pure state can be represented as:
\begin{equation}
|\psi\rangle=\sum_{k=0}^{d-1}a_k e^{i\theta_k}|k\rangle,
\label{psi}
\end{equation}
where each \( a_k \geq 0 \) is the coefficient, referred to as \textit{amplitudes}, and \( \theta_k \in [0,2\pi) \) is the \textit{phases}.

Peres considered determining unknown \( |\psi\rangle \) with the probability distributions under two bases \( \mathcal{B}_0 \) and \( \mathcal{B}_1=F\mathcal{B}_0 \), where $F$ is the quantum Fourier transform \cite{peres1997}. Using basis \( \mathcal{B}_0 \), the probability distribution \( \{|\langle \psi|k\rangle|^2\} \) provides us with knowledge of the amplitudes.
\begin{equation}
    p_k=|\langle \psi|k\rangle|^2=a_k^2
\end{equation}
Thus, \( a_k=\sqrt{p_k} \). The number \( p_k \) can be estimated by measuring \( |\psi\rangle \) large enough times and counting the frequency of outcome \( k \). When \( a_k=0 \), the term \( a_k e^{i\theta_k} \) is also zero, rendering the phase \( \theta_k \) meaningless. 
So the question is how to calculate the phases \( \theta_k \) for the nonzero amplitudes by $\{|\langle\psi|F|k\rangle|^2\}_{k=0}^{d-1}$.
For \( d=2 \), it's easy to demonstrate that there are at most two distinct sets of phases that can be computed, meaning two candidates. 
Peres's conjecture has been proven correct.

\begin{figure}[!ht] \centering\includegraphics[width=0.4\textwidth]{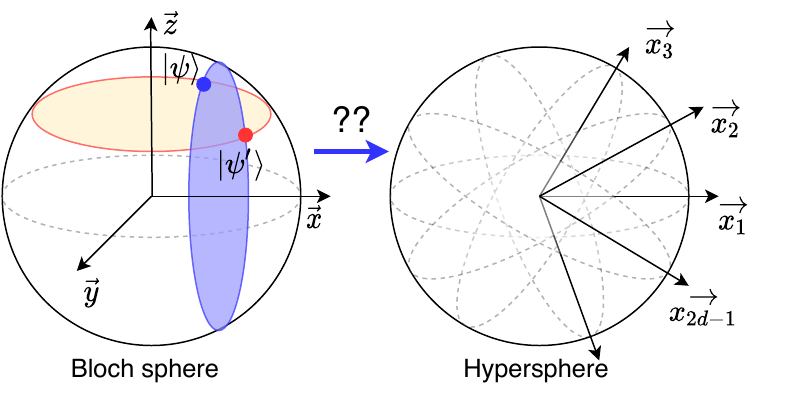}
\caption{Geometric perspective.}  
\label{geo}
\end{figure}

We provide a brief explanation with the illustration in Fig.(\ref{geo}). Up to a global phase, $d=2$, each pure qubit state corresponds one-to-one with a point on the Bloch sphere. The two complementary observables are the Pauli $X$ and $Z$ operators. Given an unknown state $|\psi\rangle$, the measurement result of Pauli $Z$ determines the red circle to which $|\psi\rangle$ belongs. Specifically, we create a plane perpendicular to the $z$-axis through the point $|\psi\rangle$, and the intersection of this plane with the Bloch sphere forms an arc. All points on this arc correspond to pure states that satisfy the probability distribution $\{|\langle \psi| k \rangle|^2\}$. The yellow points inside the circle represent collections of mixed states that match the distribution $\{|\langle \psi| k \rangle|^2\}$. 

Now we explain the red circle and its inner part. 
Actually, it is determined by 
\begin{equation}
    \mbox{tr}(\rho |0\rangle\langle 0|)= |\langle \psi|0\rangle|^2
\end{equation}
The left part is the Schmidt-Hilbert inner product of $(\rho,|0\rangle\langle 0|)$. For example, consider the plan decided by $3x+4y=5$. The left side is the inner product of vector $(x,y)$ and $(3,4)$. The vector $(3,4)$ decides the direction perpendicular to the plane and $5$ decides the specific position of the plane. We consider the Schmidt-Hilbert inner product in the Hilbert space, where $\rho$ is the state that can vary and $|0\rangle\langle0|$ is the direction. If $\mbox{tr}(\rho |0\rangle\langle 0|)=0$, the red plan is just the equator. 
 If $\mbox{tr}(\rho |0\rangle\langle 0|)=1$, the red plan is the north pole for state $|0\rangle$. 
We did not consider $\mbox{tr}(\rho |1\rangle\langle 1|)= |\langle \psi|1\rangle|^2$ as $|\langle \psi|0\rangle|^2+|\langle \psi|1\rangle|^2=1$. 

Similarly, we create a plane perpendicular to the $x$-axis through the point $|\psi\rangle$, and the intersection of this plane with the Bloch sphere forms another arc. All points on this arc correspond to pure states that satisfy the probability distribution $\{|\langle \psi| F |k \rangle|^2\}$. The number of intersection points between the two arcs is at most two. 
If point $|\psi\rangle$ is located at the endpoints of the $z$-axis or the $x$-axis, only one intersection point appears. 

All pure qudits form a real unit hypersphere of dimension $2d-1$. 
For each $|k\rangle \langle k|$ or $F|k\rangle \langle k|F^{\dag}$, it corresponds to an axis. 
Are the $2d-2$ axes for unitary $I$ and $F$ still enough to filter finite candidates? How can we choose two unitary operations to minimize the number of intersection points?

Unfortunately, when \( d=4 \), there exist infinite sets of phases that satisfy the relationship $\{|\langle\psi|F|k\rangle|^2\}$ \cite{sun2020}. And the intersection points are infinite.

Indeed, it is a phase retrieval problem \cite{balan2006signal,waldspurger2015phase,jaganathan2016phase}, to construct the minimal number of unitary operations \( \{U_1,\cdots, U_k\} \), such that \( \{|\langle \psi|U_j|l\rangle|^2:j=1,\cdots,k\} \) can guarantee unique solutions for the phases \( \theta_k \).
Flammia, Silberfarb, and Caves consider that the unknown pure state is not arbitrary but belongs to a subset where we disregard a measure-zero set from all pure states. For the states in this subset, any single unitary operation \( U_1 \), not just the Fourier transformation, fails to provide a unique determination of phases \cite{flammia2005}.
Up to now, it's evident that at least three additional unitary operations are necessary to uniquely determine the phases for \( d=3 \) and \( d\ge 5 \) \cite{carmeli2015many}.

Our first finding reveals that a single unitary operation \(U\), with all elements below the secondary diagonal set to zero and nonzero elements elsewhere, assists in narrowing down a maximum of finite \(2^{d-1}\) sets of phases to match the probability distribution \(\{|\langle \psi|U|k\rangle|^2\}\) for nearly all pure states. The instances where this approach fails form a measure-zero set of all pure states, contingent upon the specific structure of \(U\). As an example, we consider \(U\) with all real-number elements, represented as \(U=\sum_{k=0}^{d-1}|\psi_k\rangle\langle k|\).
The states \(\{|\psi_k\rangle\}\) form an orthonormal basis, thus \( U^{\dag}U=\sum_k |k\rangle\langle k|=I \).  
\begin{equation}
    \mathcal{C}_1=\{|\psi_1\rangle,\cdots,|\psi_{d-1\rangle}\}
\end{equation}
These states are normalized versions of orthogonal states \( |\tilde{\psi}_k\rangle \), given by \( |\psi_k\rangle=|\tilde{\psi}_k\rangle/|\langle \tilde{\psi}_k|\tilde{\psi}_k\rangle|^2 \). 
The unnormalized states \( |\tilde{\psi}_k\rangle \) are defined as 
\begin{equation}
|\tilde{\psi}_j\rangle =
\begin{cases}
\sum_{k=0}^{j}A_k|k\rangle-A_{j+1}|j+1\rangle, & \text{for } 0\le j\le d-2, \\
\sum_{k=0}^{d-1}A_k|k\rangle, & \text{for } j=d-1
\end{cases}
\label{psij}
\end{equation} 
Here, the coefficients \( \{A_k:k=0,\cdots,d-1\} \) are chosen to ensure the orthogonality of these states. One possible choice is given by 
\begin{equation}\label{Ak}
A_0=1, \quad A_{j+1}^2=\sum_{k=0}^{j}A_k^2, \quad \text{for } 0\le j\le d-2
\end{equation}
We may as well assume all \( A_j \) to be positive numbers. 

When \( d=2 \), the basis \( \mathcal{C}_1 =\mathcal{B}_1 \) represents the eigenbasis of the observable Pauli \( X \). The unitary operation \( U \) corresponds to the Hadamard operation, a 1-qubit Fourier transform.

For general $d$, these basis states are orthogonal.
For $0\le j<k \le d-1$, 
\begin{equation}
\langle\tilde{\psi}_j|\tilde{\psi}_k\rangle=\langle\tilde{\psi}_j|\tilde{\psi}_{j+1}\rangle=\sum_{k=0}^j A_k^2-A_{j+1}^2=0 
\end{equation}

 With the bases $\mathcal{B}_0$ and $\mathcal{C}_1$, we can obtain two probability distributions. For each probability distribution, we consider the former $d-1$ probabilities, as the last one can be expressed by 1 minus the sum of others. 
\begin{equation}
\left\{
\begin{aligned}
p_k & = |\langle \psi|k\rangle|^2 =a_k^2, \\
q_k & =|\langle \psi|U|k\rangle|^2=\frac{|\sum_{j=0}^kA_ja_je^{i\theta_j}-A_{k+1}a_{k+1}e^{i\theta_{k+1} }|^2}{\sum_{k=0}^{k+1}A_k^2} ,
\end{aligned}
\right.
\end{equation}
where $k=0,\cdots,d-2$. 
For a state \(|\psi\rangle\), both the number of independent amplitudes and phases are \(d-1\). Since \(\sum_k a_k^2=1\) and the global phase can be ignored, our task is to calculate the \(2d-2\) independent coefficients of amplitudes and phases with the \(2d-2\) equations.

Case 1:  All the amplitudes are nonzero.

In this case, we need to determine \(d-1\) phases, assuming we let the first phase \(\theta_0\) be zero. The calculation is straightforward. 
We rewrite the complex value of $\sum_{j=0}^kA_ja_je^{i\theta_j}$ as follows 
\begin{equation}\label{Sk}
    S_k e^{i\alpha_k}=\sum_{j=0}^{k}A_ja_je^{i\theta_j}
\end{equation}
Here $0\le k\le d-2$, $S_k\ge 0$ and $\alpha_k\in [0,2\pi)$. 
Now the probability $q_k  =|\langle \psi|U|k\rangle|^2$ is actually the following one. 
\begin{equation}
	\begin{split}
	q_k &= \frac{1}{\sum_{k=0}^{k+1}A_k^2} |S_k-A_{k+1}a_{k+1}e^{i(\theta_{k+1}-\alpha_k)}|^2\\
	    &= \frac{S_k^2+A_{k+1}^2 a_{k+1}^2-2S_k A_{k+1}a_{k+1}\cos(\theta_{k+1}-\alpha_k)}{\sum_{k=0}^{k+1}A_k^2}
	\end{split}
\label{qk}
\end{equation}
The unknown parameter in Eq.(\ref{qk}) is only $\theta_1$ when $k=0$. 
 After obtaining all amplitudes via basis \(\mathcal{B}_0\), the amplitudes $\{a_j\}$ are all known and the nonzero parameters $\{A_k\}$ are defined in Eq.(\ref{Ak}). 
 In this case, all $\{a_k\}$ are nonzero. So $S_0=A_0a_0\ne 0$. 
 \begin{equation}
     \cos \theta_1=\frac{(A_0a_0)^2+(A_1a_1)^2-q_0(A_0^2+A_1^2)}{2S_0A_1a_1}
 \end{equation}
 Thus at most two values of $\theta_1$ will be calculated. 

For each possible value of $\theta_1$, we can calculate the corresponding parameters $S_1$ and $\alpha_1$ by Eq.(\ref{Sk}). 
We iteratively use the Eq.(\ref{qk}) ($k=1$), at most two values of $\theta_2$ will be calculated. 
Unless we have the relation that $S_1=0$. 

We then iteratively use Eq.(\ref{Sk}) and Eq.(\ref{qk}) for $k=2,\cdots,d-2$. We can calculate the phases by the sequence as follows
\begin{equation}\label{chain}
    \theta_0=0\to \theta_1 \to \cdots\to \theta_{d-1}
\end{equation}
Thus at most $2^{d-1}$ set of phases can be calculated. 
This chain will be broken when 
\begin{equation}\label{fail1}
    S_k=\sum_{j=0}^kA_ja_je^{i\theta_j}=0, \mbox{~for some~} k\in[1,d-1].
\end{equation} 
For some \(S_k = 0\), the equation for \(q_k\) provides no information about \(\theta_{k+1}\). In this case, any value of \(\theta_{k+1}\) would satisfy the equation for \(q_k\). Consequently, there are infinitely many candidates that match the probabilities \(\{q_k\}\).

  Case 2: Certain amplitudes may be zero. 
  
  We denote the indices of non-zero amplitudes as \(\{k_0,\cdots,k_{j-1}\}\), where \(0\leq k_0<\cdots<k_{j-1}\), and \(j\leq d\). 
Then the pure state has the form
$|\psi\rangle=\sum_{l=0}^{j-1}a_{k_l}e^{i\theta_{k_l}}|k_l\rangle$. 
And we need to determine $\{\theta_{k_1},\cdots,\theta_{k_j}\}$, as $\theta_{k_0}=0$ for the freedom of global phase. 

Following a similar approach to the previous analysis, we utilize the probability \(q_{k_1-1}\) to compute up to two values of \(\theta_{k_1}\). Substituting each value of \(\theta_{k_1}\) into Eq.(\ref{qk}), we can subsequently calculate up to two values of \(\theta_{k_2}\) using \(q_{k_2-1}\). The sequential calculation of each phase no longer follows the pattern in Eq.(\ref{chain}), but proceeds as 
\begin{equation}  \label{chain2}
\theta_{k_0}=0\to \theta_{k_1} \to \cdots\to \theta_{k_{j-1}},
\end{equation} based on the probabilities \(q_{k_1-1}\) to \(q_{k_{j-1}-1}\). Overall, we can compute up to $2^{j-1}$ sets of phases to satisfy the probabilities determined by basis \(\mathcal{C}_1\). 
In this case, the chain to determine the phases will be broken if 
\begin{equation}\label{fail2}
    S_{k_l}=0, \mbox{~for some~}l=1,\cdots,j-1. 
\end{equation}

As a summary, the probability distributions under \(\mathcal{B}_0\) and \(\mathcal{C}_1\), denoted as \(\{|\langle \psi|\psi_k\rangle|^2\}\) and \(\{|\langle \psi |U|\psi_k\rangle|^2\}\) respectively, can generate a maximum of \(2^{d-1}\) candidates for \(|\psi\rangle\), unless the coefficients of \(|\psi\rangle\) satisfy Eq.(\ref{fail1}) with all nonzero amplitudes, or Eq.(\ref{fail2}) with nonzero amplitudes \(\{a_{k_0},\cdots,a_{k_{j-1}}\}\).

For \(k=1,\cdots,d-1\), if \(S_k=\sum_{j=0}^{k}A_ja_je^{i\theta_j}=0\), then the following two vectors are orthogonal.
\begin{equation}\label{Akak}
    (A_0,\cdots,A_k) \perp (a_0e^{i\theta_0},\cdots,a_ke^{i\theta_k})
\end{equation} 
The constant coefficients of \((A_0,\cdots,A_{d-1})\) are defined in Eq.(\ref{Ak}). Consequently, the vector \((A_0,\cdots,A_{k})\) forms a 1-dimensional subspace \(\mathcal{V}_{k+1}\). Its orthogonal complement subspace \(\mathcal{V}_{k+1}^{\mathrm{C}}\) is \(k\)-dimensional. The remaining coefficients \((a_{k+1}e^{i\theta_{k+1}},\cdots,a_{d-1}e^{i\theta_{d-1}})\) form a \(d-1-k\) dimensional subspace \(\mathcal{W}_{k+1}\). For pure states that satisfy \(S_k=0\), they are the unit vectors of the \(d-1\) dimensional subspace \(\mathcal{N}_k\),  
\begin{equation}
\mathcal{N}_k=\mathcal{V}_{k+1}^{\mathrm{C}}\oplus\mathcal{W}_{k+1}
\end{equation}
The states \(|\psi\rangle\) that fail to be filtered into finite candidates are the unit vectors belonging to the set 
\begin{equation}
    \mathcal{N}=\bigcup_{k=1}^{d-1} \mathcal{N}_k
\end{equation} 
 Compared to the entire \(d\) dimensional space \(\mathcal{H}_d\), \(\mathcal{N}_k\) is of measure-zero. Intuitively, a one-dimensional line is a tiny portion of a two-dimensional surface.  The finite union of measure-zero sets is still a set of measure-zero. Thus, if we randomly choose a state \(|\psi\rangle\), the probability distributions \(\{|\langle \psi|\psi_k\rangle|^2\}\) and \(\{|\langle \psi |U|\psi_k\rangle|^2\}\) can produce at most \(2^{d-1}\) estimates with probability 1, unless $|\psi\rangle \in \mathcal{N}$.

The subspace \(\mathcal{N}\) for the failure case is related to the chosen basis \(\mathcal{C}_1\), as indicated in Eq.(\ref{Akak}) for \(k=1, \cdots, d-1\). We have the flexibility to set \(\{A_k\}\) to other nonzero values or modify them as needed. For instance, by setting \(A_{j+1}^2 = \sum_{k=0}^j A_k^2\), we could allow \(A_k\) to be complex numbers rather than just positive numbers. The reconstruction methods and the analysis of the failure set remain valid under these conditions. Referring back to Eq.(\ref{psij}), we observe that by replacing the Fourier operation \(F\) with the unitary \(U\), which features zero elements below the secondary diagonal and nonzero elements elsewhere, we can effectively filter a finite number of candidates for nearly all pure states. 

From a geometric perspective, we identify $2d-2$ new axes corresponding to the unitaries $I$ and $U_1$. For each point $|\psi\rangle$ on the hypersphere in Fig.(\ref{geo}), disregarding a set of measure zero, the number of intersection points between two perpendicular planes is finite. 

\section{Sequential measurements by introducing a POVM} 

In this section, we focus on the unique determination of phases using the original Peres's two bases, \(\mathcal{B}_0\) and \(\mathcal{B}_1\), by introducing a POVM. The key objective is to uniquely determine the phases for the nonzero amplitudes. 

When all the amplitudes of \( |\psi\rangle \) in Eq.(\ref{psi}) are nonzero, various methods can uniquely determine the \( d-1 \) phases using a constant or \(\log d\) number of newly designed orthonormal bases \cite{goyeneche2015five,wang2022pure,verdeil2023pure,pereira2022scalable}. For unknown states with some zero amplitudes, the initial schemes might fail, but this can be corrected by focusing on the subspace where all amplitudes are nonzero. Specifically, the results from \(\mathcal{B}_0\) can identify the positions of the nonzero amplitudes \(\{k_0, k_1, \cdots, k_{j-1}\}\). We then change into new bases adaptively to determine the phases within the \(j\)-dimensional subspace spanned by \(\{|k_0\rangle, \cdots, |k_{j-1}\rangle\}\).
Moreover, if we do not change the bases adaptively, the initial schemes work with probability 1. The failure cases, where unknown pure states have some zero amplitudes, form another measure-zero set. Without loss of generality, let's consider the first amplitude to be zero. In this case, \( |\psi\rangle \) is a unit vector in the \(d-1\) dimensional subspace spanned by \((0, a_1e^{i\theta_1}, \cdots, a_{d-1}e^{i\theta_{d-1}})\). The positions of the nonzero amplitudes can vary, but the union of \(d\) measure-zero sets remains measure-zero. 

The measurements discussed above involve projections onto rank-1 operators \(\{|\Psi_k\rangle\langle\Psi_k|\}\). Upon reading the measurement outcome, the state \(|\psi\rangle\) will collapse into one of these states \(|\Psi_k\rangle\), resulting in the loss of all information (amplitudes and phases) about the initial state \(|\psi\rangle\). In contrast, weak measurements allow observers to gather minimal information about the system on average while causing minimal disturbance to the state \(|\psi\rangle\) \cite{ritchie1991realization,oreshkov2005weak,tamir2013introduction}. The wavefunction can be weakly measured by the position observable, followed by a strong measurement using the momentum observable. This process enables a direct readout of the complex value of \(\psi(x,t)\) at the position $x$ of interest.

We consider using a different POVM (Positive Operator-Valued Measure) to measure the state \( |\psi\rangle \) first, rather than relying on rank-1 projections or weak measurements. This approach will alter the original state, potentially aiding in the extraction of phase information without losing all the information about the initial state.

We partition all $d$-dimensional pure states into the following classes:
\begin{equation}\label{classes}
    D_{k_0,k_1,\cdots,k_{j-1}}^{j}
\end{equation}
Here $j$ means the number of nonzero amplitudes, $0<j\le d$. 
And the positions of nonzero amplitudes are $k_0,\cdots,k_{j-1}$, where $0\le k_0 <\cdots < k_{j-1}\le d-1$. There are $2^{d-1}$ classifications in total. By $\{|\langle \psi|k\rangle|^2\}$, we know the set where $|\psi\rangle$ belongs. 

It is the trivial case when $j=1$. As the state $|\psi\rangle$ is one of the computational states in $\mathcal{B}_0$. 
We consider $j=d$ and use the following POVM to measure state $|\psi\rangle$ in Eq.(\ref{psi}). 
\begin{equation}\label{2outcome}
    \{M_0=|l_1\rangle\langle l_1|+|l_2\rangle\langle l_2|, M_1=I-M_0\}
\end{equation}
We can obtain the result 0 with probability $\langle \psi|M_0|\psi\rangle$, and the collapsed state will be 
\begin{equation}
|\psi(l_1,l_2)\rangle =\frac{a_{l_1}e^{i\theta_{l_1}}|l_1\rangle + a_{l_2}e^{i\theta_{l_2}}|l_2\rangle}{a_{l_1}^2+a_{l_2}^2}  
\label{two}
\end{equation}

 Lemma: Suppose we know the nonzero amplitudes of the state in Eq.(\ref{two}). To determine the phase difference \(\theta_{l_1}-\theta_{l_2}\), we can use the following two rank-1 projectors \( |\Phi_1\rangle\langle \Phi_1| \) and \( |\Phi_2\rangle\langle \Phi_2| \):
\begin{equation}
\left\{
\begin{aligned}
|\Phi_1\rangle &= A_1|l_1\rangle + A_2 e^{i\alpha}|l_2\rangle \\
|\Phi_2\rangle &= B_1 |l_1\rangle + B_2 e^{i\beta}|l_2\rangle
\end{aligned}
\right.
\end{equation}
Here \( A_1, A_2, B_1, B_2 > 0 \), \( A_1^2 + A_2^2 \le 1 \), \( B_1^2 + B_2^2 \le 1 \), \(\alpha\) and \(\beta\) are known, and \(\sin(\alpha - \beta) \ne 0\).

Proof: Using these two projections, we obtain the following two probabilities:
\begin{equation}
\left\{
\begin{aligned}
|\langle\psi(l_1, l_2)|\Phi_1|^2 &= \frac{M + 2N \cos[(\theta_{l_2} - \theta_{l_1}) - \alpha]}{(a_{l_1}^2 + a_{l_2}^2)^2} \\
|\langle\psi(l_1, l_2)|\Phi_2|^2 &= \frac{M + 2N \cos[(\theta_{l_2} - \theta_{l_1}) - \beta]}{(a_{l_1}^2 + a_{l_2}^2)^2}
\end{aligned}
\right. \label{MN}
\end{equation}

Here \( M = a_{l_1}^2 A_1^2 + a_{l_2}^2 A_2^2 > 0 \) and \( N = 2a_{l_1}A_1 a_{l_2}A_2 > 0 \). 
In Eq.(\ref{MN}), the only unknown parameter is \(\theta_{l_2} - \theta_{l_1}\). We have the following observations:
\begin{equation}
\cos[(\theta_{l_2} - \theta_{l_1}) - \alpha] = \cos(\theta_{l_2} - \theta_{l_1}) \cos\alpha + \sin(\theta_{l_2} - \theta_{l_1}) \sin\alpha
\end{equation}

Thus, we obtain two linear equations in terms of \(\cos(\theta_{l_2} - \theta_{l_1})\) and \(\sin(\theta_{l_2} - \theta_{l_1})\) by Eq.(\ref{MN}). The unique solutions can be calculated when the determinant is nonzero. Specifically, when \(\sin(\alpha - \beta) \ne 0\), we can determine the phase difference \(\theta_{l_2} - \theta_{l_1}\) using Eq.(\ref{MN}). \(\qed\)

Now we turn to the following procedure. 
\[
  \Qcircuit @C=0.8em @R=0.8em {
\lstick{|\psi\rangle} &\qw  &\gate{\mbox{POVM}} &\gate{\mbox{Inverse Fourier}} &\meter \\
}
\]
 We use one POVM to change the state into Eq.(\ref{two}) and then project it with the basis $\mathcal{B}_1$.  
The quantum Fourier transform is the following. 
\begin{equation}
F=\frac{1}{\sqrt{d}}\left(
  \begin{array}{ccccc}
    1 & 1 & 1 & \cdots &1\\
    1 & w & w^2 & \cdots &w^{d-1}\\
    1 & w^2 & w^4 & \cdots &w^{2(d-1)}\\
    \vdots& \vdots& \vdots& & \vdots\\
    1 & w^{d-1} & w^{2(d-1)} & \cdots &w^{(d-1)(d-1)}
  \end{array}
\right),
\end{equation}
where $w=e^{\frac{2\pi i}{d}}$. 
For $k=0,\cdots,d-1$, we know 
\begin{equation}
\begin{split}
	|\langle\psi(l_1,l_2)|F|k\rangle|^2 &=\frac{1}{d} |\langle \psi(l_1,l_2)|(w^{kl_1}|l_1\rangle+w^{kl_2}|l_2\rangle) |^2\\
        &= \frac{1}{d}|\langle \psi(l_1,l_2)|(|l_1\rangle+e^{\frac{2\pi i}{d}k(l_2-l_1)}|l_2\rangle) |^2 
	\end{split}
\end{equation}
Since \( l_1, l_2 \in \{0, 1, \cdots, d-1\} \), it follows that \( l_2 - l_1 \in [1-d, d-1] \). If \( |l_2 - l_1| \ne d/2 \), then the probabilities \( |\langle \psi(l_1, l_2) | F | k_1 \rangle |^2 \) and \( |\langle \psi(l_1, l_2) | F | k_2 \rangle |^2 \) measured by the basis \(\mathcal{B}_1\) can determine the phase difference when \( |k_1 - k_2| \ne d/2 \). This is justified by the condition of the lemma above:
\begin{equation}
\sin \frac{2\pi (k_2 - k_1)(l_2 - l_1)}{d}  \left\{
\begin{aligned}
&= 0 \quad \text{if} \quad |k_1 - k_2| = d/2, \\
&\ne 0 \quad \text{otherwise}.
\end{aligned}
\right.
\end{equation}

To summarize, we can determine the phase difference \(\theta_{l_2} - \theta_{l_1}\) for the two nonzero amplitudes \( a_{l_1} \) and \( a_{l_2} \) using the following procedure, provided \( |l_1 - l_2| \ne d/2 \). First, we apply the POVM in Eq.(\ref{2outcome}) to obtain a number of states as given in Eq.(\ref{two}). Then, by measuring with the basis \(\mathcal{B}_1\) to obtain any two probabilities \( q_{k_1} \) and \( q_{k_2} \), the phase difference can be uniquely determined, as analyzed in the lemma, provided \( |k_1 - k_2| \ne d/2 \).

Consider we have known the set Eq.(\ref{classes}) where the state $|\psi\rangle$ belongs to. Now we can define a new POVM to uniquely determine all the phases like the chain in Eq.(\ref{chain2}), where $j\ge 2$.

When the dimension $d$ is odd, for any two values $k_m,k_n\in \{k_0,\cdots,k_{j-1}\}$, $k_m-k_n$ is always an integer, which is not equal to $d/2$. 
The POVM could be the following: 
\begin{equation}\label{POVM}
    \{G_0,\cdots,G_{j-2},I-\sum_{k=0}^{j-2}G_k\}
\end{equation}
Here $G_0=(|k_0\rangle\langle k_0|+|k_1\rangle\langle k_1|)/2$, $\cdots$, 
$G_{j-2}= (|k_{j-2}\rangle\langle k_{j-2}|+|k_{j-1}\rangle\langle k_{j-1}|)/2 $. 

When the dimension $d$ is even.  The strategy will not work for the states in the sets $D_{0,\frac{d}{2}}^2$, $D_{1,\frac{d}{2}+1}^2$, $\cdots$, or $D_{\frac{d}{2}-1,d-1}^2$. For the states in the other $2^d-1-d/2$ sets, we can always rearrange a sequence $k_0,\cdots,k_{j-1}$ such that the interval of the adjacent position is not equal to $d/2$. Namely, $k_{l+1}-k_l\ne d/2$ for $l=0,\cdots,j-2$. Then by the POVM in Eq.(\ref{POVM}), we can obtain $j-1$ kinds of different collapsed states and then determine the phase differences by basis $\mathcal{B}_1$. Then all the phases can be uniquely determined.

\section{Conclusion}

In this study, we discuss the determination of all finite-dimensional pure states using two observables. Without ancilla, one observable can only provide information about the amplitudes. At least one additional observable is needed to determine the phases. 
Firstly, we show that with two observables, the two probability distributions \(\{|\langle \psi|k\rangle|^2\}_{k=0}^{d-1}\) and \(\{|\langle \psi|U|k\rangle|^2\}_{k=0}^{d-1}\) can yield at most \(2^{d-1}\) distinct estimations of \(|\psi\rangle\) unless \(|\psi\rangle\) belongs to a measure-zero set.
Secondly, we demonstrate that by using an adaptive POVM followed by projecting the collapsed states onto the basis \(F\mathcal{B}_0\), all the phases can be uniquely determined. This method is effective except when the dimension \(d\) is even and \(|\psi\rangle\) contains two nonzero amplitudes at specific positions. 

 There are several interesting topics to discuss further. 
 Firstly, we designed a unitary operator \( U \) where all elements below the secondary diagonal are zero. This operator can be decomposed into the product of \( d-2 \) two-level unitary matrices. It is interesting to explore efficient experimental implementations of this decomposition or others in various physical systems.
 Secondly, when an experimental setup can randomly prepare a large number of different quantum states \(\{|\psi_1\rangle, \cdots, |\psi_N\rangle\}\), their probability distributions under two sets of projection measurements will, with high probability, differ unless the states belong to a zero-measure set or a finite set of candidate states. This implies that after repeated measurements, we will observe different experimental results. Therefore, using these two sets of observables to determine certain properties of quantum states or to distinguish between them is an interesting topic. Finally, we use a chain sequence to determine all phases. This analysis can be used in the construction of other orthonormal bases for phase determination. For instance, it is worth investigating whether a constant number of experimentally feasible orthonormal bases can be constructed to determine all pure-state phases without excluding the measure-zero set. Additionally, when \(d=2^n\), it would be interesting to explore if fewer than \(2n-1\) sets of orthogonal measurements, either with unentangled projections or with the introduction of a small number of entanglement resources, can provide a finite set of candidates or uniquely determine the $n$-qubit pure state by neglecting a measure-zero set.

\textbf{Acknowledgements---}
We would like to express our gratitude to Sixia Yu for his valuable suggestions. This work was supported by the National Natural Science Foundation of China under Grants No. 62001260 and No. 42330707, as well as the Beijing Natural Science Foundation under Grant No. Z220002.

\bibliographystyle{unsrt}   
\bibliography{main}   

\end{document}